\documentclass[referee,usenatbib]{mnras}
\usepackage{mathptmx}
\usepackage[T1]{fontenc}
\usepackage{ae,aecompl}
\usepackage{graphicx}	% Including figure files
\usepackage{amsmath}	% Advanced maths commands
\usepackage{amssymb}	% Extra maths symbols
\title[self-similarity]{Equilibrium solution for cold dynamical systems and self-similarity.}
\author[Alard, C.]{Alard, C., 
\\
IAP, 98bis Boulevard Arago, Paris \\}
\date{}
\pubyear{2016}
\begin{document}
\label{firstpage}
\pagerange{\pageref{firstpage}--\pageref{lastpage}}
\maketitle
%
% Abstract of the paper
\begin{abstract}
Numerical simulations demonstrate a link between dynamically cold initial solutions and an evolution towards self-similarity.
However the nature of this link is not fully understood. In this work the link between cold initial conditions and self-similarity near equilibrium is established. The evolution towards self-similarity is analyzed using an analytical solution in a power-law potential. The analytical solution indicates a convergence towards self-similarity after a number of dynamical times even if the inital conditions are far from self-similarity. The power-law model is extended by using perturbative analysis. The perturbative analysis shows that once the power-law potential is initiated it tends to become stronger and propagate. This behavior demonstrates the mechanism behind the convergence towards auto-similarity. The cold solutions are compatible with a broad range of self-similar solutions. As a consequence some seed of a specific self-similarity class must appear to induce a convergence mechanism. In practice some local induction of a power-law potential is necessary and some examples of such inductive mechanisms are given.
\end{abstract}
gravitation
\section{Introduction}
\label{Intro}
 The cold dark matter (CDM) paradigm in cosmology implies that structures formed from initial conditions with a very small
 velocity dispersion. High resolution numerical simulations of CDM structure formation show that a self similar regime (see \cite{Gunn}) appears in the central region of halos and extend for about two decades (see \cite{Ludlow}). The origin of this self similar regime and the convergence towards a self similar solution near equilibrium is well established for specific conjectures (see \cite{Bertschinger1985}  \cite{Binney}, \cite{Lancellotti}, \cite{Alard2013b}, \cite{Halle}, \cite{Schulz}). However a fundamental problem remains. How is it possible that starting from various non self-similar initial conditions the system evolves after a number of dynamical times towards self-similarity ?
Furthermore since a large class of self-similar solutions exists, how does the system makes a choice and evolves towards a specific
similarity index ? Here are the question we will try to answer in this work.
\section{Late time properties of dynamically cold solution.}
After a long dynamical time the central region of a cold solution will develop a very large number of very close
 folds in phase space (see \cite{Fillmore}). This ensemble of very close folds converges towards the continuous solution at the center. If the system
is finite this central region has to reach a stationary state, thus converging towards the limit of a  stationary continuous
solution at the center. In this central limit the stationary solution is a function of the energy.
\subsection{the initial conditions.}
Cold initial conditions in the 2D phase space $(x,v)$ can be represented by an horizontal line with an infinitely small thickness $\zeta$.
 The density profile of the line  in the $x$ direction is $\rho(x)$. The coordinate system is defined in order to have 
the same mass on each side of the origin. As a consequence the force $F=-\frac{d \phi}{d x}$
is zero at the origin in this coordinate system.
\label{sub_1}
\subsection{Velocity scaling.}
By applying a scaling transformation in the velocity space to an horizontal line in
in phase space we still obtain an horizontal line. The thickness $\zeta$ is changed by the velocity re-scaling, but since
we are in the limit $\zeta \to 0$ the line itself is un-changed. The only quantity changed in the initial conditions due to the velocity re-scaling is the total mass. As a consequence a re-scaling in velocity should produce a final state identical 
to the un-scaled solution except for a re-scaling of the total mass. Note that a re-scaling of the total mass also implies a re-scaling of the potential. Let now apply a re-scaling in velocity to a stationary solution $f(x,v)$. This solution must be a function of the energy. 
\begin{equation}
 f(x,v)=F\left(E\right)=F\left(\phi(x)+\frac{v^2}{2} \right)
 \label{Eq_f} 
\end{equation} 
The transform in velocity affects the total mass and as a consequence the potential $\phi(x)$. Considering
the transformed quantities $v_2$ and $\phi_2$ we should observe
the following transformation $f_2$ for $f$,
\begin{equation}
 \begin{cases}
  v_2= \left(1+\epsilon \delta v \right) v \\
  \phi_2=\phi+\epsilon \delta \phi \\
  f2=(1+\epsilon \delta k) f
 \end{cases}
 \label{Eq_scale}
\end{equation}
Note that in Eq. (\ref{Eq_scale}) the re-scaling of $f$ is associated to the re-scaling of the total mass. Let's now introduce
the transformations defined in Eq. (\ref{Eq_scale}) in Eq. (\ref{Eq_f}) and develop the calculations. We will consider
the case $\epsilon \ll 1$ and develop Eq. (\ref{Eq_f}) to the first order in $\epsilon$.
\begin{equation}
 f_2 \simeq F\left(\phi(x)+\frac{v^2}{2} \right)+F^{'} \left(\phi(x)+\frac{v^2}{2} \right) \left(\delta \phi + \delta v v^2 \right) \epsilon
\label{Eq_e}
\end{equation}
Using Eq. (\ref{Eq_scale}) and Eq. (\ref{Eq_e}), equating the terms in $\epsilon$, and using the
variables $(x,E)$ we find,
\begin{equation}
 \delta \phi(x) -2 \delta v \phi(x)= -2 \delta v E + \delta k \frac{F(E)}{F^{'}(E)}
\label{Eq_e2}
\end{equation}
The left side of (\ref{Eq_e2}) depends on $x$ while the right side depends on E, since $x$ and $E$ are 
 two independent variables, both sides of the equation have to be equal to a constant. As a consequence
we obtain 2 equations,
\begin{equation}
\begin{cases}
 \delta \phi(x) =2 \delta v \phi(x)+c_0 \\
 \delta k \frac{F(E)}{F^{'}(E)}=2 \delta v E-c_0
\end{cases} 
\label{Eq_e3}
\end{equation}
 The solution of Eq. (\ref{Eq_e3}) for the variable $E$ leads to,
\begin{equation}
 \begin{cases}
 F(E)=c_1 \left(\phi(x)+\frac{v^2}{2} + \tilde c_0)^{\alpha} \right) \\
 \alpha=\frac{\delta k}{2 \delta v}
 \end{cases}
 \label{Eq_e4}
\end{equation}
Here $\tilde c_0=-\frac{c_0}{2 \delta v}$.
If we take a small area of size $\zeta$ around the origin in the initial conditions, since the
force is zero at the origin (see Sec. \ref{sub_1}) this area will remain at the origin at equilibrium.
 This small area represent the continuous
limit of the smooth equilibrium solution at the origin.
 Since for the cold solution we have the limit $\zeta \to 0$, the density
profile corresponds to a delta function, and thus the density at the center is infinite. Assuming that
the exponent in Eq. (\ref{Eq_e3}) is negative this requires that,
\begin{equation}
 \begin{cases}
 \lim_{x \to 0} \ \ \ \tilde {\phi}(x)=0 \\
  \tilde{\phi}(x)=\phi(x)+\tilde c_0 
 \end{cases}
\label{Eq_force0}
\end{equation}
\subsection{Consequence for the potential.}
\label{Sec_pot}
By integrating the phase space density in Eq. (\ref{Eq_e4}) we obtain the density and using the Poisson Equation (Eq. \ref{Eq_poisson}) we find,
\begin{equation}
 \frac{d^2 \tilde{\phi}}{d x^2}-c_2 \tilde{\phi}^{\alpha+\frac{1}{2}}=0
 \label{Eq_pot}
\end{equation}
Where $\frac{c_2}{c_1}$ is a function of $\alpha$. After performing an integration of Eq. (\ref{Eq_pot}) we obtain,
\begin{equation}
 \begin{cases}
 \left(\frac{d \tilde{\phi}}{d x}\right)^2-c_2 \tilde{\phi}^{\alpha+\frac{3}{2}} +c_3=0 \\
  k_2=\frac{4 k}{2 \alpha+3}
 \label{Eq_pot2}
 \end{cases}
\end{equation}
The left side of Eq. (\ref{Eq_pot2}) $\frac{d \phi}{d x}$ is the force. At the origin the force is
zero (see Sec. \ref{sub_1}), while the other term must be also zero at the origin (see Eq. \ref{Eq_force0}), thus
we must have $c_3=0$.
A general solution of Eq. (\ref{Eq_pot2}) can be obtained by introducing the new variable $u(x)$. Using this new
variable Eq. (\ref{Eq_pot2}) transforms to,
\begin{equation}
\begin{cases}
 \beta^2 \left(\frac{d u(x)}{d x}\right)^2 -c_2 \\
 u(x)=\tilde{\phi}(x)^{\beta} \\
 \beta=\frac{4}{1-2\alpha}
\end{cases}
\label{Eq_pot3}
\end{equation}
The solution to Eq. (\ref{Eq_pot3}) is straightforward leading to,
\begin{equation}
\begin{cases}
 u(x)=\frac{\sqrt{c_2}}{\beta} x + c_4 \\
 \tilde{\phi}(x)=\left(\frac{\sqrt{c_2}}{\beta} x + c_4 \right)^{\frac{1}{\beta}}
\end{cases}
\label{Eq_pot4}
\end{equation}
According to Eq. (\ref{Eq_force0}) we must have $\tilde{\phi}(0)=0$, using Eq. (\ref{Eq_pot4}) we obtain
$\tilde{\phi}(0)=(c_4)^{\frac{1}{\beta}}$. Eq. (\ref{Eq_pot3}) indicates that for $\alpha<0$ we have $\beta>0$,
as a consequence we must have $c_4=0$. Using Eq. (\ref{Eq_e4}) the final expression for the phase space density at equilibrium in the central region reads,
\begin{equation}
 f(x,v)=c_1 \left( \left(\sqrt{c_5} x \right)^{\frac{1}{\beta}} + \frac{v^2}{2} \right)^{\alpha}
\label{Eq_f_self}
\end{equation}
It is important to note that in Eq. (\ref{Eq_f_self}) $\alpha$ and $c_5$ are free parameters. As a consequence this general
asymptotic self-similar solution in the central region is consistent with a large class of self-similar solutions. The choice of a particular solution will have to be induced by a specific mechanism. Such mechanism will drive the solution towards a specific value of $\alpha$.
%
%%%%%%%%%%%%%%%%%%%%%%%%%%%%%%%%%%%%%%%%%%%%%%%%%%%%%%%%%%%%%%%%%%%%%%%%%%%%%%%%%%%%%%%%%%%%%%%%%%%%%%%%%%
%                                                                                                        %
%%%%%%%%%%%%%%%%%%%%%%%%%%%%%%%%%%%%%%%%%%%%%%%%%%%%%%%%%%%%%%%%%%%%%%%%%%%%%%%%%%%%%%%%%%%%%%%%%%%%%%%%%%
%
%
\section{Solution of the Vlasov equation in 1D for a power law potential.}
\label{Pw_sec}
In the former section it was demonstrated that the phase space density at equilibrium converges towards a power law solution
near the origin. This power-law behavior is associated with self-similar solutions. The problem is that there should be some general mechanism leading to this self-similar equilibrium solution. Starting from general non self-similar cold initial conditions, we will now investigate the converge towards auto-similarity.
For a phase space density $f(x,v,t)$ in one spatial dimension $x$ and associated velocity dimension $v$ the Vlasov equation reads:
\begin{equation}
   \frac{\partial f}{\partial t}+\frac{\partial f}{\partial x} v- \frac{\partial \phi}{\partial x} \frac{\partial f}{\partial t} = 0
\label{Eq_vlasov}
\end{equation}
The function $\phi=\phi(x,t)$ is the potential related to f by the Poisson equation,
\begin{equation}
  \begin{aligned}
    \Delta \phi & = \rho \\
    \rho & = \int f dv
  \end{aligned}
\label{Eq_poisson}
\end{equation}
Here we are interested in finding an analytical solution for a power law potential. Note that
a power law potential does not imply that the solution in this potential is self similar. To derive the solution 
we will apply the method used in (\cite{Alard2013a}). However instead of applying the method to
the self similar equation we will apply the method directly to the Vlasov equation. In this approach
we will study the evolution of the phase space density in a constant power-law potential and for general
cold initial conditions. We thus consider,
\begin{equation}
  \phi(x) = k x^{\beta+2}
 \label{Eq_pot}
\end{equation}
Let now transform Eq. (\ref{Eq_vlasov}), by introducing Eq. (\ref{Eq_pot}),and
the following change in variable, 
\begin{equation}
\begin{cases}
   \begin{aligned}
    u&=&|x|^\eta \\
 \eta&=&\frac{\beta}{2}+1\\
  \end{aligned}
  \end{cases}
  \label{Eq_pot}
\end{equation}
We also make appropriate choices for the scales of x and v in order to have $k=\eta$. Then Eq. (\ref{Eq_vlasov}) reads,
\begin{equation}
  \frac{\partial G}{\partial t}+sgn(x) \left( \eta v \frac{\partial G}{\partial u} u^{\frac{\eta-1}{\eta}}-\eta \frac{\partial G}{\partial v} u^{\frac{2 \eta-1}{\eta}} \right)=0
 \label{Eq_vlasov2}
\end{equation}
Here we define,
$$
 G(u,v,t)=F(x,v,t)
 $$
 and,
 $$
 sgn(x)=
 \begin{cases}
   \begin{aligned}
    -1  \ \ \ \  & x <0 \\
    1  \ \ \ \  & x \geq 0
   \end{aligned}
  \end{cases}
 $$
 We introduce again a new change in variables in Eq. (\ref{Eq_vlasov2}),
 \begin{equation}
   \begin{aligned}
     \begin{cases}
     R &= \sqrt{u^2+v^2} \\
     \psi &= \arccos\left(\frac{u}{R} \right)
     \end{cases}
   \end{aligned}
   \label{Eq_R_phi}
 \end{equation}
 With this new change of variables Eq. (\ref{Eq_vlasov2}) now reads,
 \begin{equation}
   \frac{\partial H}{\partial t}-\eta R^{\kappa}|\cos(\psi)|^{\kappa}\frac{\partial H}{\partial \psi}=0
   \label{New_Eq}
 \end{equation}  
 Here we define,
 \begin{equation}
 \begin{cases}
   H(R,\psi,t)=F(x,v,t) \\
   \kappa=\frac{\eta-1}{\eta}
 \end{cases}
 \label{Eq_kappa}
 \end{equation}
 The Vlasov equation in the new variables (\ref{New_Eq}) has a general solution,
 \begin{equation}
   \begin{cases}
     f(x,v,t)=F(R,Q) \\
     Q= t R^{\kappa}  +\frac{1}{\eta} \int |\cos(\psi)|^{-\kappa} d\psi 
   \end{cases}
  \label{Eq_Sol}
 \end{equation}
 \subsection{Solution for a dynamically cold system.}
 We are interested in a dynamically cold solution of the Vlasov equation. Thus
 Eq. (\ref{Eq_Sol}) must correspond to a spiral with infinitely small thickness in the $(R,\psi)$ space. 
 For points out of this spiral the solution has a zero value. 
 The solution in Eq. (\ref{Eq_Sol}) is a function of the two variables $(R,Q)$, on the spiral itself the
 internal equation of the spiral implies a relation between these two variables.
 \begin{equation}
   Q=G(R)
 \label{Eq_Q}
 \end{equation}
  Now let use the equation of the spiral (Eq. \ref{Eq_Q}) to redefine the density in phase space using the new variables
  $\frac{Q}{G(R)}$,  
  \begin{equation}
     F(R,Q)=H \left(R,\frac{Q}{G(R)}\right) \\
   \label{Eq_sol_cold}
  \end{equation}
 On the spiral Eq. (\ref{Eq_sol_cold}) indicates that $f(x,v,t)=H(R,1)$ while out of the spiral we have $f(x,v,t)=H(R,C)$ with
 $C \neq 1$. Consequently the general cold solution is,
  \begin{equation}
   \begin{cases}
     f(x,v,t)=H\left(R,\frac{Q}{G(R)}\right) \\
     H(R,1)=P(R) \\
     H(R,C)=0 \ \ \rm if \ \ C \neq 1
   \end{cases}
   \label{Eq_sol_cold2}
  \end{equation}
We see that the solution in Eq. (\ref{Eq_sol_cold}) satisfies all the requirements of the cold solution. This solution is zero out of the spiral and has a value dependent of the value of the pseudo distance in phase space $R$ on the spiral.
\subsection{Evolution towards self-similarity.}
\label{Sub_sec_self}
By combining Eq's (\ref{Eq_Q}) and (\ref{Eq_Sol}) an equation for the spiral
is obtained,
\begin{equation}
  t R^{\kappa}  -\int |\cos(\psi)|^{\kappa} d\psi=P(R)
  \label{Eq_spiral0}
\end{equation}
 To estimate the behavior of Eq. (\ref{Eq_spiral0}) at late time and near the center of the system ($R \to 0$)
 we need to consider the two following cases. First the case (I), $P(R)$ dominates the term in $R^{\kappa}$ when $R \to 0$. 
The second case (II) is when $P(R)$ is of the same order or weaker than the term in $R^{\kappa}$.
It is important to note that in general $P(R)$ is not a power law and does not correspond to a self similar spiral.

The first case is easy to analyze, if the behavior is dominated by $P(R)$ Eq. (\ref{Eq_spiral0})  does not depend on time in 
the regime $R \to 0$. Consequently we are left with a stationary solution for small value of $R$, and this case is not of interest.
If we now consider case (II), it is clear that at late time the term in $R^{\kappa}$
due to its co-factor in time tends to dominate the other term in $R$, $P(R)$.  As a consequence for small values of $R$ or large
values of the time $t$, Eq. (\ref{Eq_spiral0}) is reduced to,
\begin{equation}
t R^{\kappa}  -\int |\cos(\psi)|^{\kappa} d\psi=0
\label{Eq_spiral}
\end{equation}
The inter fold distance for the associated spiral is obtained by considering
a variation of of $\phi$ to $\phi + 2 \pi$ and $R$ to $R+\delta R$ in
Eq. (\ref{Eq_spiral}). The variation in $\phi$ leads to a constant term,
resulting in the following formula for inter-fold distance $\delta R$,
\begin{equation}
\frac{\delta R}{R} \propto \frac{R^{-\kappa}}{t}
\label{Eq_fold_dist}
\end{equation}
It is interesting to note that Eq. (\ref{Eq_fold_dist}) is very similar to Eq.
(21) in (\cite{Alard2013a}),
\begin{equation}
\frac{\delta R}{R} \propto R_0^{\frac{1}{\alpha_2}}
\label{Eq_alard2013}
\end{equation}
  The pseudo phase space distance $R_0$ in Eq. (\ref{Eq_alard2013}) is a self similar quantity, thus $R_0=\frac{R}{t^{\alpha_2}}$,
  and $\beta \alpha_1=-2$ (see \cite{Alard2013a} Sec. 5). By using also Eq. (\ref{Eq_kappa}), then
  Eq. (21)  in (\cite{Alard2013a}) reduces exactly to Eq. (\ref{Eq_fold_dist}). This proves that
  the two spirals in phase space are identical, and that as a consequence the late time evolution
  of the solution for a power-law potential is self-similar. It is clear that the reduction of Eq. (\ref{Eq_spiral0}) to Eq. (\ref{Eq_spiral}) is an effect of the late time evolution due to the time co-factor
in Eq. (\ref{Eq_spiral0}). Early in the evolution of the system or in the initial conditions the term $P(R)$
is present but is erased by the the late time evolution of the system. Basically the non-self similar initial
conditions are diluted by the infinite number of turns of the solution near the center.
\subsection{Decomposition of the solution.}
\label{Sec_decomp}
At some late stage of its dynamical evolution the cold solution will approach a stationary state which should be close to an equilibrium solution when smoothed.
For a power law potential corresponding to a density, $\rho=x^\beta$, the density
in phase space for a smooth equilibrium solution must be of the form,
\begin{equation}
\begin{cases}
  g(x,v)=R^\gamma \\
  \gamma=\frac{\beta-2}{\beta+2}
\end{cases}
\label{Eq_smooth}
\end{equation}
The appropriate smoothing to apply to the cold solution in order to obtain a smooth
equilibrium solution is adaptive. The scale of this adaptive smoothing should follow
the spiral inter distance between two consecutive folds. To apply this type of smoothing
to the cold solution let first decompose the solution in a stationary but spatially
dependent part and a time time dependent but spatially constant part. To perform this
decomposition one has to notice that in Eq. (\ref{Eq_sol_cold}) the value of the density
on the spiral depends on the variable $R$. As a consequence by dividing the phase space density by the function $P(R)$ in Eq. (\ref{Eq_sol_cold}) a  spiral with constant density on its folds is obtained. As a consequence it is possible to write
the solution as the product of a smooth function of $R$ and a spiral with constant density on its fold. The decomposition reads,
\begin{equation}
 f(x,v)=P(R) \ S(x,v)
 \label{Eq_decomp}
\end{equation}
Here $S(x,v)$ describes the spiral with constant density on its folds.
\subsection{Proportionality between the inter-fold distance and the fold thickness.}
\label{Sec_thickness}
Let us now point to an important property of the spiral in phase space, the vanishing thickness of the spiral is proportional to the inter-fold distance. This property is a direct consequence
of the fact that the spiral thickness and the inter fold distance are derived from the same
equation with just a change in a parameter. To be more specific, the interfold distance
is the distance obtained after a variation of the angular variable $\phi$ to $\phi + 2 \pi$. 
Similarly the thickness is a consequence of slightly different initial conditions. Let consider
that one side of the spiral corresponds to a position $(R, \phi)$ in the initial condition and that
the other edge corresponds to $(R, \phi+\delta\phi)$. As a consequence the spiral thickness is the distance corresponding to a variation $\delta\phi$. In the late stage of the dynamical evolution the inter-fold is small and the thickness even smaller, with the consequence that the calculation of the distance can be linearized. Consequently, the ratio between the thickness and the inter-fold distance
will scale like $\frac{\delta\phi}{2 \pi}$ which is a constant.
\subsection{Satisfying the Poisson equation.}
\label{Sec_poisson}
It is simple to realize that
a spiral with constant density on its folds and with a thickness proportional to the inter-fold distance has constant mean density in phase space. Basically within a fold the total
mass scales like the inter fold volume, resulting in a constant averaged density inside the fold. This property is conserved in a change of coordinates provided that the coordinate change
is linear at the scale of the inter-fold distance. In this case the total mass in a fold
and the inter-fold volume are both transformed by a multiplication with the Jacobian of the coordinate change and as a consequence their ratio is conserved. We have already seen that in Sec.
(\ref{Sec_decomp}) and Eq. (\ref{Eq_decomp}) that the solution can be de-composed in a spiral with constant density on its
fold and and spatially smooth part with density $P(R)$ (see Eq. \ref{Eq_sol_cold} for a definition of $P(R)$) . Since the mean density of the smoothed
spiral is constant in phase space, the mean density in phase space is directly $P(R)$. An a consequence if $P(R)$ corresponds to the phase space density in Eq. (\ref{Eq_smooth}) then
the solution is consistent with the Poisson equation.
%
%%%%%%%%%%%%%%%%%%%%%%%%%%%%%%%%%%%%%%%%%%%%%%%%%%%%%%%%%%%%%%%%%%%%%%%%%%%%%%%%%%%%%%%%%%%%%%%%%%%%%%%%%
%
\section{The route towards self similarity.}
%consider a perturbation of the potential and the effect it has on the spiral in phase space.
We have just seen in Sec. (\ref{Sub_sec_self}) that non self-similar initial conditions in a power law potential evolves towards self-similarity at late times. We will now consider the more general case where the potential
is not a power law. More specifically we will investigate the case where the potential at the center is a power-law and not a power-law
at larger distances. In this model it is expected that self-similarity will be induced by the power-law potential at the center. Note that such a model implies that a stable power-law potential exists in a small area near the center of the system.
The origin of this potential can be some fluctuation in the density, but it must persist for at least a few dynamical times.
In the case where some similarity class is imposed by an external process, like for instance the secondary infall (see \cite{Bertschinger1985}), the only
fluctuations able to exists for a few dynamical times are those which are consistent with the dynamical forcing by this external process. In the specific case of the secondary infall there are 2 solutions for the slope of the power-law at the center,
the first corresponds to a full power-law solution, while the second is associated with a NFW like profile (see \cite{Dehnen}, \cite{Navarro}). The index of the power-law at the center can also be imposed directly by a specific process (see \cite{Alard2013a}).  
In the continuation we will illustrate how this model leads to a convergence towards auto-similarity in the central region of  the system. 
\subsection{The model for the potential.}
 The potential in the system is a power law in the vicinity of the center and any function at larger distances. It is assumed
that a local forcing of the potential exists at small scale and that it dominates the contribution to the potential due to the projected density of the system. With a proper choice of the boundaries this slightly perturbed power-law model for the potential is valid
in a small area near the center. The specific extent of this area is determined by the requirement that the perturbation
is a small fraction of the background power-law potential. As a consequence near the center we have a power-law potential $\phi_0$ plus a small perturbation due to the projected density $\delta \phi$. The total potential $\phi$ then reads,
\begin{equation}
 \phi=\phi_0+\epsilon \delta \phi
 \label{Eq_phi_pert}
\end{equation}
Where $\epsilon \ll 1$ and $\delta \phi$ is the potential associated with the total projected density of the system.
\subsubsection{Perturbed power-law near the center of the system.}
\label{Sec_pw1}
Let us define first the un-perturbed model and its un-perturbed smooth phase space density $F_0$. For the initial value of $F_0$ we take directly the initial value of the full space density of the system $F_I$. The un-perturbed model evolves in the power-law potential $\phi_0$.
The perturbed solution evolves under the influence ot the total potential $\phi$ (see Eq. \ref{Eq_phi_pert}).
The dynamics of the perturbed cold solution is modified by $\delta \phi$ leading to a perturbation of the equation
of the cold spiral in phase space. This small potential perturbation of the spiral equation will result in a small
perturbation of the associated smooth density $F$.
As a consequence the perturbed smooth phase-space density $F$ reads,
\begin{equation}
 F=F_0 + \epsilon \delta F
\label{Eq_F_pw}
\end{equation}
Where $\epsilon \ll 1$ and $\delta F$ is the perturbed phase-space density. 
\subsubsection{Convergence towards auto-similarity in a small area.}
\label{Sec_pw}
\begin{figure}
\includegraphics[width=8cm]{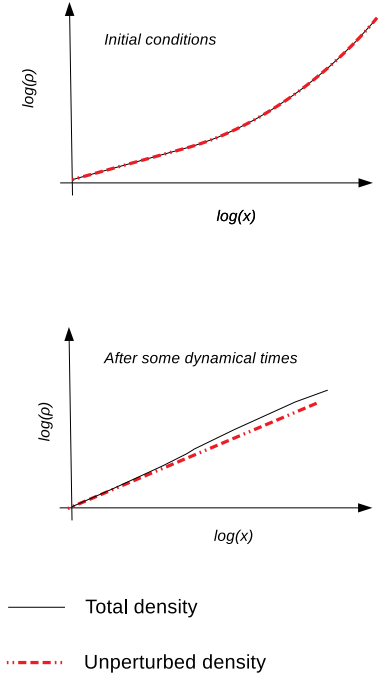}
\caption{An illustration of the convergence towards auto-similarity for a perturbed power-law potential. Initially the system
is not self-similar, and the initial density is not a power-law. The first order perturbation to the potential is induced by the system density. As a consequence the initial first order potential perturbation is not self-similar. The initial un-perturbed density corresponds to the initial density of the full system. After some dynamical times the un-perturbed density evolves towards auto-similarity. The first order perturbation to the potential corresponds to the un-perturbed density. The late un-perturbed density is self-similar and is a power-law identical to the un-perturbed potential. Consequently to the first order the potential is a power-law and we are left with only a second order perturbation to the power-law potential. The same reasoning can be re-iterated leading to full convergence.}
\label{fig1}
\end{figure}

Initially $F$ is not self similar, thus the projected density and the potential $\delta \phi$ are not power-laws.
After a number of dynamical times the un-perturbed density $F_0$
evolves towards self-similarity (see Sec. \ref{Sub_sec_self}), leading to a power-law un-perturbed projected density and potential. 
 This final un-perturbed density is a power-law identical to $\phi_0$ except for a scaling. 
Consequently at late dynamical time we find that the potential perturbation has two
parts, $\delta \phi= \delta \phi_0 + \delta \phi_1$. The first part, $\delta \phi_0$ corresponds to the contribution of the un-perturbed solution. The second part $\delta \phi_1$ corresponds to the contribution
of the perturbation which is of order $\epsilon$, thus we re-write $\delta \phi_1=\epsilon \delta \psi$. 
Using Eq. (\ref{Eq_phi_pert}). We find that the first order perturbation to the potential at late time is $\delta \phi_0$ which is a power law. Thus at first order the potential is a full power-law after a number of dynamical times. The remaining non-power-law part of the potential $\delta \psi$ is of second order in $\epsilon$.
The same method can be re-iterated with this second order perturbation by just making a substitution of $\epsilon$
for $\epsilon^2$. This process will lead to a further reduction of order of the perturbation and to a convergence towards auto-similarity after a series of loops of this type. An illustration of this process is provided in Fig. (\ref{fig1}).
\begin{figure}
\includegraphics[width=10cm]{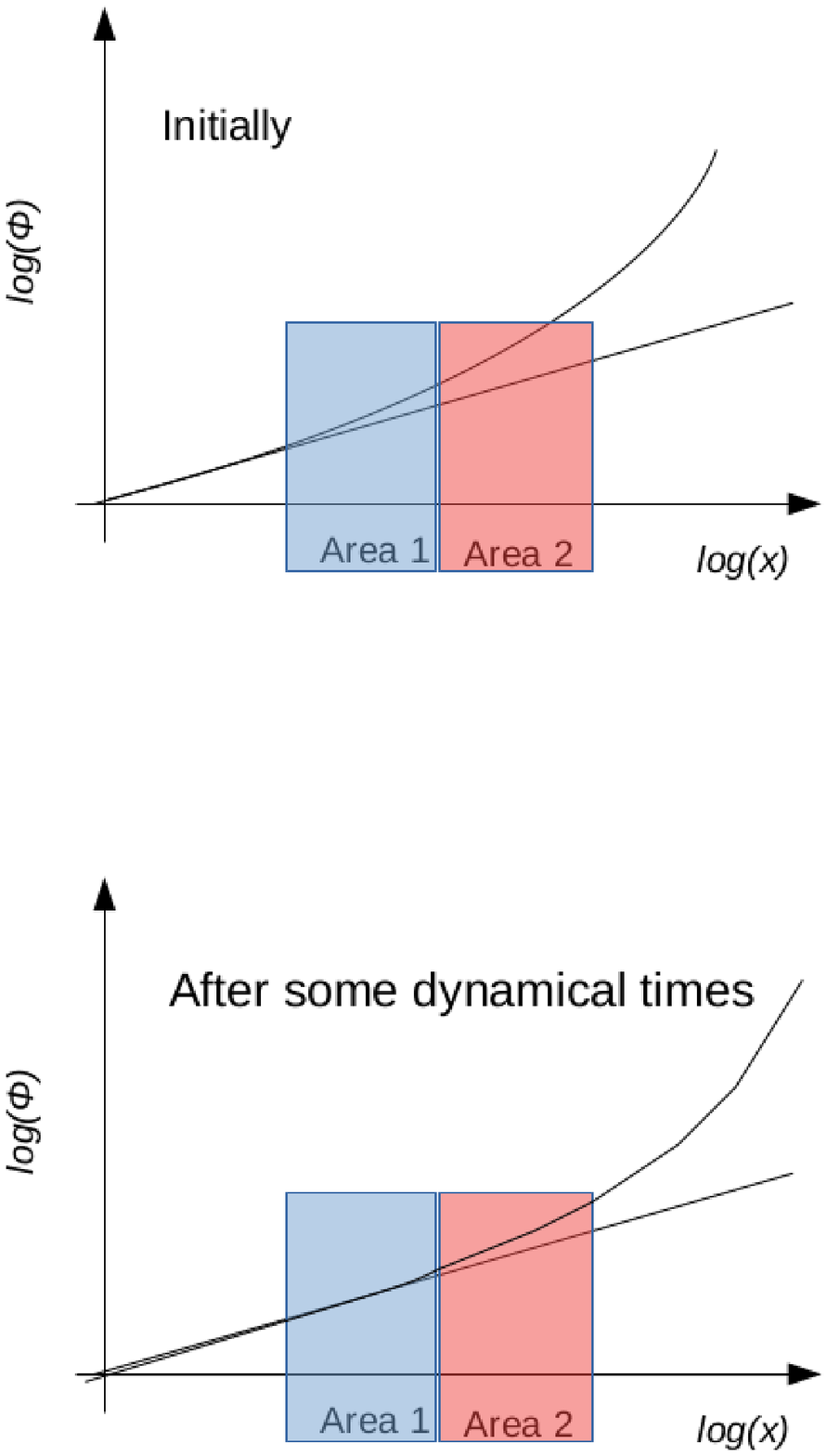}
\caption{Here we describe the process of propagation of auto-similarity from area 1 to area 2. Initially area 1 corresponds
to a first order perturbation of the power-law potential. The perturbation in area 2 is larger. The process of convergence towards auto-similarity described in Sec. (\ref{Sec_pw}) and Fig. (\ref{fig1}) operates in a few dynamical cycles. As result the potential in area 1 become a power-law. By continuity
the adjacent edge of area 2 is also asymptotically a power-law. Provided that we chose an appropriate boundary for area 2 we are again
in the situation of a perturbed power-law potential. By re-applying the former convergence process (see \ref{Sec_pw} and Fig. \ref{fig1}) to area 2 we have propagated auto-similarity to area 2.}
\label{fig2}
\end{figure}
\subsubsection{Propagation of auto-similarity.}
Let suppose that for instance we have a dominant forcing by a power-law potential near the center of the system,
but that at larger distance the self gravity of the system dominates and induce non power-law terms. Around the center of the system the potential is effectively a power-law plus a small perturbation. It was shown
in Sec. (\ref{Sec_pw}) that the system evolves towards self-similarity and the potential towards a power-law.
Now we will consider a point just outside the region where the potential is a power-law plus a small perturbation.
 Due to the evolution of the potential inside the trajectory of the point towards a power-law (see Sec. \ref{Sec_pw}), the overall potential
felt by the point is closer to a power-law.
After this evolution if we extend the area were we had a near power-law
behavior of the potential just enough to be consistent with the model of a power-law potential plus a small perturbation, we will be again
in the situation described in Sec. (\ref{Sec_pw}). Obviously the method can
be iterated, leading to a propagation of the auto-similar solution. An illustration of this propagation process is provided in Fig. (\ref{fig2}).
\subsection{The evolution towards a specific auto-similar solution.}
The dynamically cold initial conditions evolves towards an auto-similar equilibrium solution with a power-law potential (see Sec. (\ref{Sec_pot}). However the exponent of the power-law potential is a free parameter. It is clear that the specific choice of a given similarity class which is equivalent to the choice of the power-law
exponent must be induced by some process. One such process was described in Sec. (\ref{Sec_pw}) and a practical
example with a forcing of self-similarity due to the potential induced by the caustics is described in (\cite{Alard2013a}). Another example related to a forcing in the central region due to the angular momentum is described in (\cite{Alard2013b}) and (\cite{Halle2019}). It is interesting to note that the appearance of
a given similarity may not be due to an induction by a force field but by the dynamical properties of the
solution in its outer limits. One such example is the self similarity induced by the infall of cold dark matter
(\cite{Bertschinger1985}). Here the principle is the same than with the force field, the self-similarity
is dominant in some area where the convergence towards self similarity occurs and then propagates. 
\section{Multi-dimensional extension.}
 It is possible to extend the one dimensional solution presented in this work to several dimensions. The method proposed in
 (\cite{Alard2013a}, Sec. 7) for the self-similar solution can be applied directly to the solution in a power-law potential developed in this work. The other steps like the perturbative analysis and the propagation of the solution also follows
 naturally. It is also interesting to note that the decomposition of the solution (see Sec. \ref{Sec_decomp}) can be easily extended to several dimension. In effect the division by a smooth function of the energy or other variables will lead
 also to spiral with constant density on its fold. The argument on the proportionality of fold distance and fold thickness also holds in several dimension since it is due to a local linearization. 
\section{Conclusion.}
It was demonstrated in Sec. (\ref{Intro}) that cold initial conditions leads to a self similar solution near equilibrium.
By using an analytical solution for a power-law potential (see Sec. \ref{Pw_sec}) an analyze of the convergence mechanism
from dynamically cold initial conditions towards a self-similar solution near equilibrium was constructed. However the similarity index
of the equilibrium solution is a free parameter, and the choice of specific solution must be induced by some mechanism. In this view
some specific influence must drive the final state towards a specific self-similarity. The inductive power law potential acts as an attractor and induce the convergence towards self-similarity after a sufficient number of dynamical times. 
%
%%%%%%%%%%%%%%%%%%%%%%%%%%%%%%%%%%%%%%%%%%%%%%%%%%%%%%%%%%%%%%%%%%%%%%%%%%%%%%%%%%%%%%%%%%%%%%%%%%%%%%%%%
%
\section*{Data Availability}
No datasets were generated or analyzed during the current study.

\begin{thebibliography}{}
\bibitem[\protect\citeauthoryear{Alard}{2013a}]{Alard2013a}
Alard, C., 2013, MNRAS, 428, 340
\bibitem[\protect\citeauthoryear{Alard}{2013b}]{Alard2013b}
Alard, C., 2013, International workshop Vlasov-Poisson : the numerical approach and its limits, IHP, 2013
\bibitem[\protect\citeauthoryear{Binney}{2004}]{Binney}
Binney J., 2004, MNRAS, 350, 939
\bibitem[\protect\citeauthoryear{Bertschinger}{1985}]{Bertschinger1985}
Bertschinger, E., 1985, ApJS, 58, 39B 
\bibitem[\protect\citeauthoryear{Colombi}{2017}]{Colombi}
Colombi, S., Alard, C., 2017, JPlPh, 83, 7002
\bibitem[\protect\citeauthoryear{Dehnen \& McLaughlin}{2005}]{Dehnen}
Dehnen, W., McLaughlin, D., 2005, MNRAS, 363, 1057
\bibitem[\protect\citeauthoryear{Fillmore \& Goldreich}{1984}]{Fillmore}
Fillmore J. A., Goldreich P., 1984, ApJ, 281, 9
\bibitem[\protect\citeauthoryear{Gunn}{1977}]{Gunn}
Gunn J. E., 1977, ApJ, 218, 592
\bibitem[\protect\citeauthoryear{Halle, Colombi and Peirani}{2019}]{Halle}
Halle, A., Colombi, S., Peirani, S., 2019, A\&A, 621, 8
\bibitem[\protect\citeauthoryear{Lancellotti \& Kiessling}{2001}]{Lancellotti}
Lancellotti C., Kiessling M., 2001, ApJ, 549, L93
\bibitem[\protect\citeauthoryear{Ludlow {\it etal.}}{2001}]{Ludlow}
Ludlow A., Navarro J., Springel V., Vogelsberger M., Wang J., White S.,Jenkins A., Frenk C., 2010, MNRAS, 406, 137
\bibitem[\protect\citeauthoryear{Navarro, Frenk \& White}{1997}]{Navarro}
Navarro, J., Frenk, C., White, S., 1997, ApJ, 490, 493
\bibitem[\protect\citeauthoryear{Schulz {\it etal.}}{2013}]{Schulz}
Schulz, A., Dehnen, W., Jungman, G., Tremaine, S.,2013, MNRAS,431,49S
\end{thebibliography}
\end{document}